\documentclass[12pt,onecolumn,draftclsnofoot]{IEEEtran}
\usepackage{changebar}
\usepackage{cite}
\usepackage{citesort}
\usepackage{amsmath}
\usepackage{rotating}
\usepackage{times}
\usepackage{psfrag}
\usepackage{color}
\usepackage{amsfonts}
\usepackage{amssymb}
\usepackage[english]{babel}
\begin{document}
\bibliographystyle{IEEEtran}
\hyphenation{lists}
\newtheorem{theorem}{Theorem}
\newtheorem{proposition}{Proposition}
\makeatletter
\def\ifundefined{\@ifundefined}
\makeatother \setcounter{page}{1}

\title{Compression and Combining Based on Channel Shortening and Rank Reduction Techniques for Cooperative Wireless Sensor Networks}
\author{\large Qasim Zeeshan Ahmed, Ki-Hong Park, Mohamed-Slim Alouini and Sonia Aissa\\
\thanks{This work was supported by KAUST Global Cooperative Research (GCR) fund and SABIC post-doctoral fellowship.}
\thanks{Q.~Z. Ahmed, K.-H. Park, and M.-S. Alouini are with the Computer, Electrical, and Mathematical Science and Engineering (CEMSE) Division, King Abdullah University of Science and Technology (KAUST), Thuwal, Makkah Province, Kingdom of Saudi Arabia
(Email: \{qasim.ahmed,kihong.park,slim.alouini\}@kaust.edu.sa).}
\thanks{S. Aissa is with INRS, University of Quebec, Montreal, QC, Canada
(Email: sonia.aissa@ieee.org).}}

\maketitle

\vspace{-1cm}
\begin{abstract}
This paper investigates and compares the performance of wireless sensor networks where sensors operate on the principles of cooperative communications. We consider a scenario where the source transmits signals to the destination with the help of $L$ sensors. As the destination has the capacity of processing only $U$ out of these $L$ signals, the strongest $U$ signals are selected while the remaining $(L-U)$ signals are suppressed. A preprocessing block similar to channel-shortening is proposed in this contribution. However, this preprocessing block employs a rank-reduction technique instead of channel-shortening. By employing this preprocessing, we are able to decrease the computational complexity of the system without affecting the bit error rate (BER) performance. From our simulations, it can be shown that these schemes outperform the channel-shortening schemes in terms of computational complexity.
In addition, the proposed schemes have a superior BER performance as compared to channel-shortening schemes when sensors employ fixed gain amplification. However, for sensors which employ variable gain amplification, a tradeoff exists in terms of BER performance between the channel-shortening and these schemes. These schemes outperform channel-shortening scheme for lower signal-to-noise ratio.

%In order to decrease the complexity of the system channel shortening and three different kinds of reduced-rank MMSE detectors are investigated for WSNs. These detectors operate on the principles of principal components analysis (PCA), cross-spectral metric (CSM) and Taylor polynomial approximation (TPA), respectively. From our simulations it can be observed that these detectors are capable of achieving a similar bit error rate (BER) performance as the full-rank MMSE detector with significantly lower complexity. They also outperform the best sensor selection scheme in terms of BER performance when using fixed amplification factor. However, for variable gain amplification factor a tradeoff between the diversity gain and the receiver complexity can be observed.
\end{abstract}
%%%%%%%%

\section{Introduction}
In wireless sensor networks (WSNs), the fundamental task is to broadcast data from the origin sensor to the destination. However, due to the limited size, power and cost of these sensors, a low power signal is often transmitted to the destination~\cite{Sayed-2007,Krishna-2008,Hou-2005}. This low power signal is further attenuated due to the propagation loss. To combat this problem, the signal is sometimes measured by as many sensors as possible~\cite{Sayed-2007,Krishna-2008}. These sensors form a distributed cooperative sensor network, enabling them to achieve spatial diversity which will help combat fading effects and extend network coverage~\cite{Laneman-2004}.

Low-complexity cooperative diversity protocols have been developed and analyzed for cooperative communications in different operating conditions and environments. According to~\cite{Laneman-2004}, the family of fixed relaying arrangements have the lowest complexity as compared to all the other families. The family of fixed relaying consists of decode-and-forward (DF) and amplify-and-forward (AF) protocols. It has been  proved that the AF protocol has the ability to achieve similar bit error rate (BER) performance as compared to that of the DF protocol, while maintaining a lower complexity~\cite{Laneman-2004,Zhao-2007}. Therefore, only the AF protocol is considered in our contribution.

The design of low-complexity detectors at the destination plays a significant role in WSNs, as the sensor nodes are powered by batteries~\cite{Sayed-2007,Krishna-2008,Hou-2005,Chen-2005}. The maximum likelihood (ML) detector is the optimal detector in terms of BER for equally likely symbols~\cite{book:Verdu}. However, due to the high computational complexity of the ML detector, suboptimal linear detectors are often considered for WSNs~\cite{Honig2000,Wang-2009,Wang-2011}. In suboptimal linear detectors, minimum mean square error (MMSE) detection is preferred due to its improved BER performance~\cite{Honig2000}. It can be observed that, as the number of sensors increases, the complexity of the MMSE detector becomes extremely high~\cite{Wang-2011}. Recently, in order to solve this problem, channel shortening (CS)-based technique has been proposed for cooperative networks~\cite{Imtiaz-2012,Imtiaz-2010}. A preprocessing matrix in CS-based technique is designed where only $U$ sensors are chosen out of $L$ sensors. The idea is to maximize the energy reception of the selected $U$ sensors while minimizing the energy leakage of the remaining sensors and the ambient noise power. As only $U$ sensors are selected and processed at the destination, the computational complexity will be lower than the ideal MMSE detector. In cooperative communications, the best relay is selected and then the transmit power of that relay is maximized~\cite{Zhao-2007,Bletsas-2006}. There are two major problems when applying this approach to WSNs. Firstly, no power adaptation is applied. Secondly, the sensors are powered by batteries, therefore, it is not rational to transmit the signal with more power. In such scenarios, CS-based techniques can be adopted and they outperform the technique of best relay selection in terms of BER as shown in~\cite{Imtiaz-2010,Imtiaz-2012}. By employing CS-based techniques, the destination captures $U$ strongest signals out of the $L$ received ones. As the receiver tries to maximize the energy of $U$ sensors, the energy in $L-U$ sensors is lost. Therefore, a loss in the BER is observed when comparing the CS-based techniques with the approach involving all participating sensors.

%\subsection{Contributions}
%The $U$ selected sensors should be available consecutively as mentioned in~\cite{Imtiaz-2010,Imtiaz-2012}. In an actual communication environment, it is possible that the best $U$ selected sensors are not consecutively placed. Therefore, in this contribution, we arrange the sensors according to SNR ordering. The main advantage of SNR ordering is that the best $U$ sensors are consecutively placed and are the first $U$ sensors. From the simulation results, it can be observed that a gain of more than $1$~dB can be achieved by using this method as compared to the CS-based approach in~\cite{Imtiaz-2010,Imtiaz-2012} when $U\geq 2$.

In order to solve this problem, a design of the preprocessing matrix with the assistance of reduced-rank techniques is proposed in this paper. Reduced-rank techniques have been widely applied to array processing~\cite{Goldstein-1997}, radar signal processing, direct-sequence code divison multiple access (DS-CDMA)~\cite{Honig2000}, space-time coded space-division~\cite{Yang-2006} and ultra-wide band (UWB) systems~\cite{Qasim-2010,Qasim-2011}. Specifically, in this contribution three types of rank reduction techniques are considered, which derive their detection subspaces based on the concepts of principal component (PC)~\cite{Qasim-2010,Alouini-2000,book:Dunteman,Honig2000}, cross-spectral metric (CSM)~\cite{Goldstein-1997,Goldstein-1997a,Qasim-2010} and Taylor polynomial approximation (TPA)~\cite{Burykh-2002,Qasim-2010}.

The main contributions of the paper are summarized as follows: (i) Arrangement of the sensors according to signal-to-noise ratio (SNR) ordering, as an improvement over CS-based techniques; (ii) Application of reduced-rank techniques for providing the compromise between computational complexity and performance; and (iii) Development of a diversity-order analysis for reduced-rank techniques.

The remainder of the paper is organized as follows. In Section~\ref{sec-sys}, a detailed explanation of the WSN model and the basic assumptions are presented. Section~\ref{sec-detector} investigates the ideal MMSE detectors for a WSN environment. It also highlights the issues of implementing an ideal MMSE detector for WSNs. Section~\ref{sec-Transformation-matrix} discusses the implementation of the preprocessing matrix. The design of the preprocessing matrix with the assistance of CS-based technique is discussed in Section~\ref{sec-Transform-matrixCS}, followed by the design of the preprocessing matrix with the assistance of reduced-rank techniques in Section~\ref{sec-Transform-matrixRR}. The complexity of all these algorithms and the ideal MMSE detector is derived and compared in Section~\ref{sec-Complexity}. Simulation results are presented in Section~\ref{sec-results}. Finally, Section~\ref{sec-summary} concludes the paper with summarizing comments.

Throughout the paper, the following notations are used. Upper case and lower case boldfaces are used for matrices and vectors, respectively. Given a matrix $\mathbf{A}$, symbols $\mathbf{A}^{*}$, $\mathbf{A}^{T}$, $\mathbf{A}^{H}$ and $\mathbf{A}^{-1}$ denote the complex conjugate, transpose, Hermitian transpose and inverse of $\mathbf{A}$, respectively.

\section{WSN System Model}\label{sec-sys}
\begin{figure}[!hbt]
\centering
\includegraphics[width=1.0\linewidth]{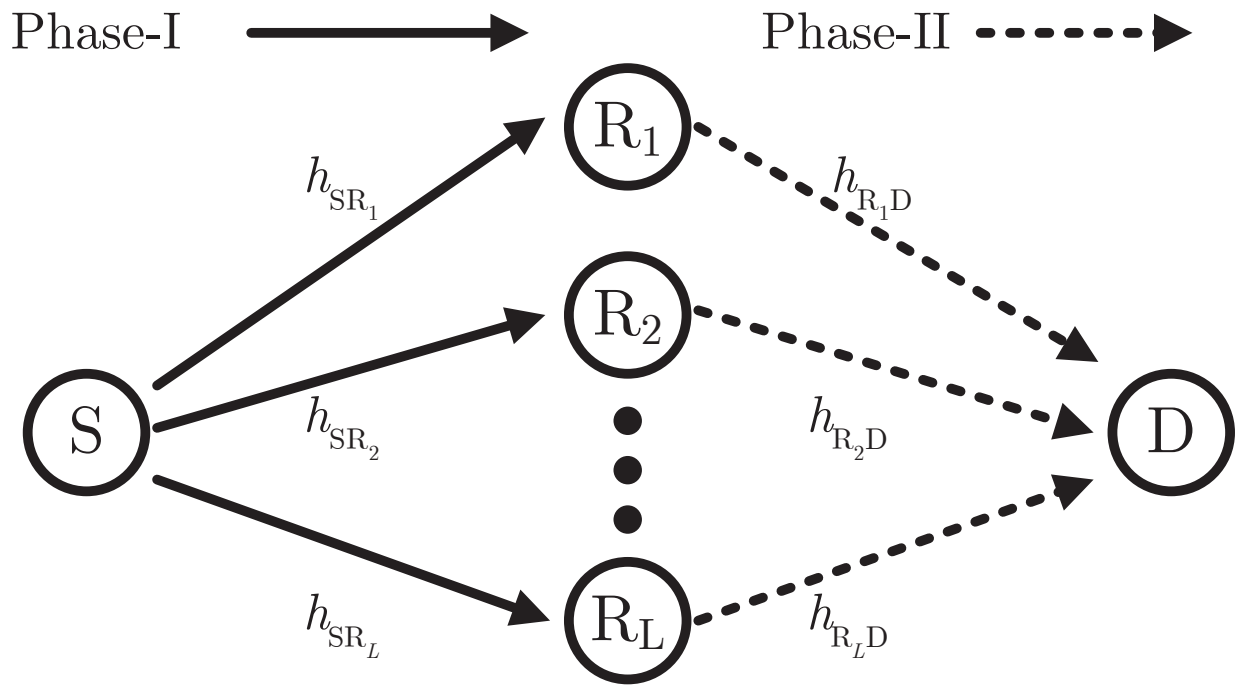}%,height=0.6\linewidth]{pic13.eps}
\caption{Schematic block diagram of a WSN when communicating with the assistance of $L$ sensors using amplify-and-forward (AF) principles.}
\label{fig:CC-1}
\end{figure}

The basic WSN system model considered in this work is shown in Fig.~\ref{fig:CC-1}. As illustrated, the source sensor $S$ transmits data to the destination $D$ with the assistance of $L$ sensors. The $L$ sensors operate on the principle of cooperative communications, such that each one amplifies and forwards the data to the destination. It is assumed that there exists no direct link between the source sensor and the destination. The channel gains for the links between the source and the $l$th sensor and from the $l$th sensor to the destination are denoted as $h_{SR_l}$ and $h_{R_lD}$, and are assumed to be mutually independent and follow the Rayleigh fading distribution model with variances $\sigma_{SR_l}^2$ and $\sigma_{R_lD}^2$, respectively. The data transmission takes place in two phases as shown in Fig.~\ref{fig:CC-1}. $S$ transmits the signal to the sensors in phase-I, while the signal is amplified and forwarded to $D$ through the intermediate sensors in phase-II. In order to minimize the interference between the sensors, orthogonality is achieved in the frequency or time domains~\cite{Bletsas-2006,Patel-2007,Zhu-2007,Chen-2004}.

\subsection{Phase-I: Transmission From Source Sensor}~\label{phase-1}
The sensor $S$ broadcasts a data symbol $b$ to all the sensors, $R_1, R_2,\cdots, R_L$. The received signals can be represented as
\begin{eqnarray}~\label{eq-1}
y_{R_l}=\sqrt{E_S}h_{SR_l}b+n_{R_l},\quad l=1,2,\cdots,L,
\end{eqnarray}
where $E_{S}$ is the average signal energy transmitted by the source and $n_{R_l}$ is a complex additive white Gaussian noise (AWGN) with mean zero and variance $\sigma_{R_l}^2$. The variance of the data symbol is assumed to be $\sigma_b^2=1$.
\subsection{Phase-II: Transmission From Relay Sensor to Destination}~\label{phase-2}
During this phase, the $l$th sensor amplifies the received signal $y_{R_l}$ by $\zeta_{R_l}$ and forwards the resulting signal to the destination. At node $D$, the signal received from the $l$th sensor is given by
\begin{eqnarray}~\label{eq-2}
y_{l}&=& h_{R_lD}\zeta_{R_l} y_{R_l}+n_{D_l}, \nonumber\\
&=&\sqrt{E_S}\zeta_{R_l}h_{R_lD}h_{SR_l}b+\zeta_{R_l}h_{R_lD}n_{R_l}+n_{D_l}.
\end{eqnarray}
Depending upon the type of sensors~\cite{Patel-2007}, the amplifying factor $\zeta_{R_l}$ can be either
\begin{eqnarray}\label{eq-3}
\zeta_{R_l}=\sqrt{\frac{E_{R_l}}{E_{S}\sigma_{SR_l}^2+\sigma_{R_l}^2}}
\end{eqnarray}
or
\begin{eqnarray}\label{eq-4}
\zeta_{R_l}=\sqrt{\frac{E_{R_l}}{E_{S}|h_{SR_l}|^2+\sigma_{R_l}^2}},
\end{eqnarray}
where $E_{R_l}$ is the average signal energy at the $l$th sensor. (\ref{eq-3}) and (\ref{eq-4}) are called as fixed-gain and variable-gain amplification factors, respectively. In the fixed-gain amplification factor, the sensor ensures that the average or long-term power constraint is maintained, but allows the instantaneous transmit power to be much larger than the average~\cite{Patel-2007,Zhu-2007,Chen-2004}. However, in the variable-gain amplification factor, each sensor uses the channel state information (CSI) from the source-sensor link to ensure that an average output energy per symbol is maintained for each realization~\cite{Patel-2007,Zhu-2007,Chen-2004}. This operation is performed at all the sensors.
%%%%%%%%%%%%%%%%%%%%%%%%%
\subsection{Receiver Structure}
As the desired signal $b$ arrives at the destination with the assistance of $L$ sensors, $L$ copies of the desired signal need to be collected. The vector form of the received signal can be represented by

%%%%%%
\begin{eqnarray}~\label{eq-5}
\pmb{y}&=&\pmb{h}b+\pmb{n},
\end{eqnarray}
%%%%%%
where the channel and noise vectors, $\pmb{h}$ and ${\pmb{n}}$, are defined as follows:
\begin{eqnarray}\label{eq-6-7}
\pmb{h}&=&\left[\sqrt{E_S}\zeta_{R_1}h_{R_1D}h_{SR_1},\cdots,\sqrt{E_S}\zeta_{R_L}h_{R_LD}h_{SR_L}\right]^T\nonumber\\&=&[h_1,h_2,\cdots,h_L]^T,\\
\pmb{n}&=&\left[\zeta_{R_1}h_{R_1D}n_{R_1}+n_{D_1},\cdots,\zeta_{R_L}h_{R_LD}n_{R_L}+n_{D_L}\right]^T\nonumber\\&=&[n_{1},n_2,\cdots,n_{L}]^T.
\end{eqnarray}
If the channel knowledge is available, the noise part can be approximated as complex Gaussian noise with zero mean and variance given by
\begin{eqnarray}\label{eq-8}
\sigma_l^2=\zeta_{R_l}^2|h_{R_lD}|^2\sigma_{R_l}^2+\sigma_{D}^2, \quad l=1,2,\cdots, L.
\end{eqnarray}
Therefore, $\pmb{n}$ is a complex Gaussian with mean zero and variance $\pmb{\Sigma}$. The variance $\pmb{\Sigma}$ will be a diagonal matrix of size $L$ and can be expressed as
\begin{eqnarray}~\label{eq-9}
\pmb{\Sigma}= \textrm{diag} [\sigma_1^2,\sigma_2^2,\cdots,\sigma_L^2].
\end{eqnarray}
%%%%%%%%%%%
%
\section{MMSE Detection for WSNs}\label{sec-detector}
%%%%%%%%%%%

\begin{figure}[!hbt]
\centering
\includegraphics[width=1.0\linewidth]{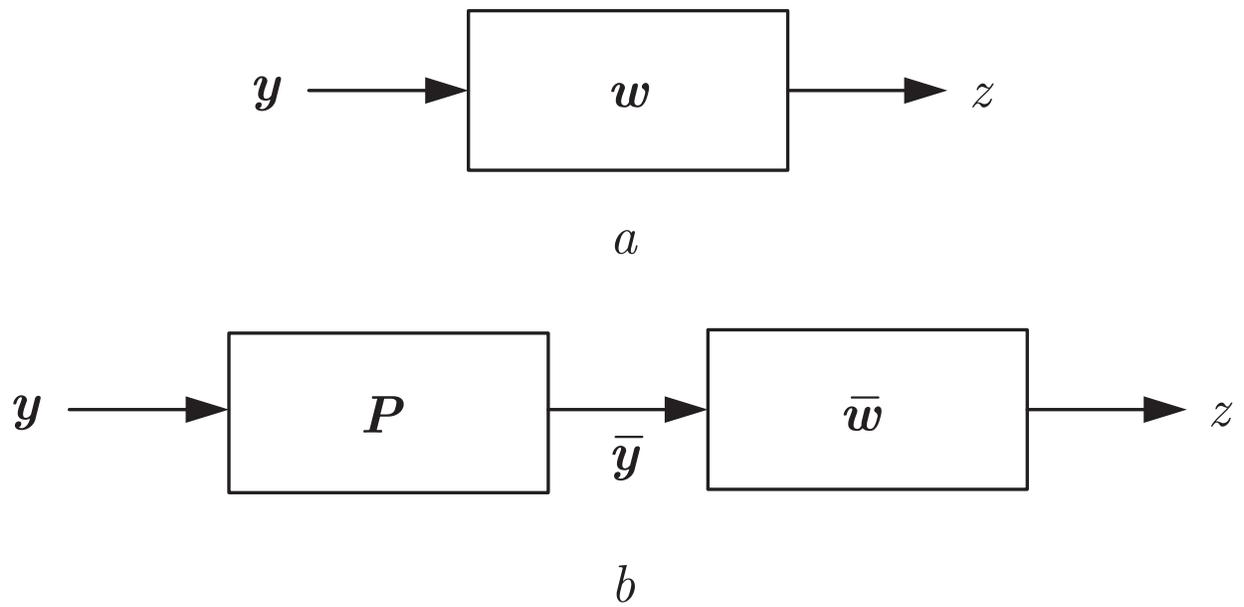}%,height=0.6\linewidth]{pic13.eps}
\caption{Block diagram of an MMSE detector for WSN when communicating with the assistance of $L$ sensors using AF principles where a) is the ideal MMSE detector and b) is the MMSE detector with the preprocessing matrix $\pmb{P}$. }
\label{fig:receiver}
\end{figure}

The receiver schematic block diagram for the ideal MMSE receiver is shown in Fig.~\ref{fig:receiver}a. To estimate the desired data bit, the receiver consists of a linear filter characterized by
\begin{eqnarray}~\label{eq-10}
z&=&\pmb{w}^H\pmb{y}=\pmb{w}^H\pmb{h}b+\pmb{w}^H\pmb{n},
\end{eqnarray}
where $\pmb{w}=[w_1,w_2\cdots,w_L]^T$ and $w_l$ is the $l$th tap coefficient of complex valued filter. The linear detector minimizes the mean-square error (MSE) cost function, i.e.,
\begin{eqnarray}~\label{eq-11}
\hspace{-0.35in}
J(\pmb{w})&=&E[|b-\pmb{w}^H\pmb{y}|^2]\nonumber\\
&=&1-\pmb{w}^HE[b^*\pmb{y}]-E[\pmb{y}^Hb]\pmb{w}+\pmb{w}^HE[\pmb{y}\pmb{y}^H]\pmb{w},
\end{eqnarray}
where $E[\cdot]$ represents the expected operator. The optimal weights for an MMSE detector can be easily obtained by derivation of (\ref{eq-11}) with respect to $\pmb{w}$ and setting to zero. The optimal weights can be easily determined as~\cite{book:Haykin}
\begin{eqnarray}~\label{eq-12}
\pmb{w}&=&\pmb{R}^{-1}\pmb{\rho},
\end{eqnarray}
where $\pmb{\rho}=E[\pmb{y}b^*]=\pmb{h}$ is the cross-correlation vector between $\pmb{y}$ and $b^*$ and $\pmb{R}=E[\pmb{y}\pmb{y}^H]=\pmb{h}\pmb{h}^H+\pmb{\Sigma}$ is the
auto-correlation of $\pmb{y}$. By substituting (\ref{eq-12}) in (\ref{eq-11}), the cost function can be expressed as
\begin{eqnarray}\label{eq-13}
{{J}}=1-\pmb{\rho}^H\pmb{R}^{-1}\pmb{\rho}.
\end{eqnarray}
From (\ref{eq-13}), it can be observed that in order to minimize the MSE we need to maximize the $\pmb{\rho}^H\pmb{R}^{-1}\pmb{\rho}$ which corresponds to maximizing the power of $z$. It can be observed from (\ref{eq-12}) that the complexity of the ideal MMSE detector is determined by the inverse of $\pmb{R}$ which is $(L\times L)$ dimensional matrix. Inverting a matrix of this size requires a computational complexity of $\mathcal{O}(L^3)$. In WSNs, the size of $L$ is usually very large, therefore, the complexity of the ideal MMSE detector will be extremely high. If the length of $\pmb{y}$ is reduced to $U$, where $U \ll L$, then the computational complexity can be reduced significantly. Therefore, in order to reduce the complexity of the ideal MMSE detector, a preprocessing matrix $\pmb{P}$ is designed in the upcoming section.
%%%%%%%%%%%%%%%%%%%%%
\section{Design of The Preprocessing matrix}~\label{sec-Transformation-matrix}
%%%%%%%%%%%%%%%%%%%%%%
The receiver block diagram for the preprocessing matrix $\pmb{P}$ is shown in Fig.~\ref{fig:receiver}b. The design of the preprocessing matrix will operate in two modes. In the first mode, a preprocessing matrix $\pmb{P}$ is designed so that the received data which is of length $L$ gets reduced to $U$, where $U<L$. Therefore, for a received vector $\pmb{y}$, the $U$-dimensional received vector is now given by
%The aim of designing the preprocessing matrix is to process $U$ signals out of $L$ signals such that the loss in BER performance is minimal.
%
\begin{eqnarray}\label{eq-14}
\bar{\pmb{y}}=\pmb{P}^H\pmb{y},
\end{eqnarray}
where $(\;\bar{\cdot}\;)$ indicates that the vector is now reduced to size $U$ instead of $L$. In the second mode, this $\bar{\pmb{y}}$ is passed through a $U$ dimensional filter. The modified cost function can now be given as
\begin{eqnarray}\label{eq-15}
J(\bar{\pmb{w}})&=&E[|b-\bar{\pmb{w}}^H\bar{\pmb{y}}|^2].
\end{eqnarray}
Similarly, as mentioned in (\ref{eq-11}), the optimal weight vector can be given as
\begin{eqnarray}\label{eq-16}
\bar{\pmb{w}}=\bar{\pmb{R}}^{-1}\bar{\pmb{\rho}},
\end{eqnarray}
where $\bar{\pmb{R}}$ is the autocorrelation matrix of $\bar{\pmb{y}}$, which is reduced to a $(U\times U)$ matrix as compared to a $(L\times L)$ matrix. This reduced-complexity scheme
requires a computational complexity of $\mathcal{O}(U^3)$ to determine the inverse of $\bar{\pmb{R}}$. As $U < L$, the complexity of the proposed system will be significantly lower than that of the ideal MMSE detector having complexity of $\mathcal{O}(L^3)$. Let us now design an optimal or an efficient preprocessing matrix $\pmb{P}$ with the assistance of CS-based technique.
%%%%%%%%%%%%%%%%%%%%%%%%%%%%%
\section{Preprocessing matrix Through Channel Shortening}\label{sec-Transform-matrixCS}
%%%%%%%%%%%%%%%%%%%%%%%%%%%%%
As the CS-based techniques work differently for time and frequency orthogonal channels, we revisit them separately as mentioned in~\cite{Imtiaz-2010,Imtiaz-2012}.
\subsection{Time Orthogonality}\label{sec-TOCS}
For time orthogonality, we assume that there is a preprocessing vector $\pmb{p}$ such that
\begin{eqnarray}\label{eq-17}
\pmb{p}=[p_1, p_2, \cdots ,p_U]^T,
\end{eqnarray}
where $U$ is the length of the filter. The received signal $\pmb{y}$ will be convolved with $\pmb{p}$ to generate the output out of which $U$ is selected to be processed by the reduced optimal weight vector $\bar{\pmb{w}}$. The output of the convolution can be written as
\begin{eqnarray}\label{eq-18}
\pmb{a}=\pmb{y}*\pmb{p}=(\pmb{A}b+\pmb{N})\pmb{p},
\end{eqnarray}
where $\pmb{A}$ and $\pmb{N}$ are the convolution matrices of $\pmb{h}$ and $\pmb{n}$, with dimension $(L+U-1)\times U$. The $``*"$ sign in (\ref{eq-18}) represents convolution. The size of the output vector $\pmb{a}$ is $(L+U-1)$. As we can only process $U$ elements of $\pmb{a}$, we require $U$ elements to be non-zero and the other $(L-1)$ elements to be zero ideally. The location of these $U$ non-zero elements may be anywhere within $\pmb{a}$, but for simpler processing they should be consecutively placed:
\begin{eqnarray}\label{eq-19}
\pmb{a}=[0, 0, \cdots, 0, a_i, a_{i+1}, \cdots, a_{i+U-1}, 0, 0, \cdots, 0]^T,
\end{eqnarray}
where $i$ is an arbitrary number such that $i=1,2,\cdots,L$. The channel-shortened received signal will now be represented as
\begin{eqnarray}\label{eq-20}
\bar{\pmb{y}}= [a_i,a_{i+1},\cdots, a_{i+U-1}]^T.
\end{eqnarray}
$\pmb{A}$ in (18) will now consist of two sub-matrices $\pmb{A}_U$ and $\pmb{A}_{L-1}$, where the selected signals will be placed in $\pmb{A}_U$ and the signals to be compressed will be placed in $\pmb{A}_{L-1}$. The energy of the selected $U$ branches is given by $\pmb{p}^H\pmb{A}_U^H\pmb{A}_U\pmb{p}$, while the energy of the remaining $(L-1)$ branches is given by $\pmb{p}^H\pmb{A}_{L-1}^H\pmb{A}_{L-1}\pmb{p}$ and the noise energy is $\pmb{p}^HE[\pmb{N}^H\pmb{N}]\pmb{p}$. We require an optimum value of $\pmb{p}$, which maximizes the energy of the selected branches and minimizes the energy of noise and the $(L-1)$ remaining branches. We can reduce the above problem to a Rayleigh quotient, by placing a constraint on the energy of the $(L-1)$ remaining branches and the noise such that $\pmb{p}^H(\pmb{A}_{L-1}^H\pmb{A}_{L-1}+E[\pmb{N}^H\pmb{N}])\pmb{p}=1$, then
\begin{eqnarray}\label{eq-21}
\pmb{p}=\arg\max_{\pmb{p}} \frac{\pmb{p}^H(\pmb{A}_U^H\pmb{A}_U)\pmb{p}}{\pmb{p}^H( \pmb{A}_{L-1}^H\pmb{A}_{L-1}+E[\pmb{N}^H\pmb{N}])\pmb{p}}.
\end{eqnarray}
Since this is a basic Rayleigh quotient problem, a well known solution is mentioned in~\cite{Imtiaz-2012,Melsa-1996,Martin-2005-1,Martin-2005}. Letting $\pmb{C}=\pmb{A}_U^H\pmb{A}_U$ and $\pmb{B}=\pmb{A}_{L-1}^H\pmb{A}_{L-1}+E[\pmb{N}^H\pmb{N}]$, the optimal value of $\pmb{p}$ can be evaluated as
\begin{eqnarray}\label{eq-22}
\pmb{p}=\pmb{F}^{-1}\pmb{v},
\end{eqnarray}
where $\pmb{F}$ is the Cholesky factor of $\pmb{B}$, such that $\pmb{B}=\pmb{F}^H\pmb{F}$, and $\pmb{v}$ is the eigenvector corresponding to the maximum eigenvalue of $(\pmb{F}^H)^{-1}\pmb{C}(\pmb{F})^{-1}$.
\subsection{Frequency Orthogonality}\label{sec-FOCS}
Frequency orthogonality has been recently utilized instead of time orthogonality among the channels because of better BER performance as shown in~\cite{Imtiaz-2010,Imtiaz-2012}. The received signal can be represented as
\begin{eqnarray}\label{eq-23}
\pmb{y}=\pmb{D}\pmb{b}+\pmb{n}=\pmb{d}+\pmb{n},
\end{eqnarray}
where
\begin{eqnarray}\label{eq-24}
\pmb{D}\!\!\!\!&=&\!\!\!\!\textrm{diag} \left[\sqrt{E_S}\zeta_{R_1}h_{R_1D}h_{SR_1},\cdots,\sqrt{E_S}\zeta_{R_L}h_{R_LD}h_{SR_L}\right],\nonumber\\
\pmb{b}\!\!\!\!&=&\!\!\!\![b, b ,\cdots, b]^T,\\
\pmb{d}\!\!\!\!&=&\!\!\!\!\left[\sqrt{E_S}\zeta_{R_1}h_{R_1D}h_{SR_1}b,\cdots,\sqrt{E_S}\zeta_{R_L}h_{R_LD}h_{SR_L}b\right]^T\nonumber.
\end{eqnarray}
Let $\pmb{p}=[p_1, p_2, \cdots, p_L]^T$ be the processing vector of size $L$, then the output signal processed through $\pmb{p}$ can be given as
\begin{eqnarray}\label{eq-25}
\pmb{p}^H\pmb{y}=\pmb{p}^H\pmb{d}+\pmb{p}^H\pmb{n}.
\end{eqnarray}
As only $U$ signals are required from $L$ in order to reduce the complexity, $\pmb{d}$ can be defined as
\begin{eqnarray}\label{eq-26}
\pmb{d}=\pmb{d}_U+\pmb{d}_{L-U},
\end{eqnarray}
where $\pmb{d}_U$ will consist of the required $U$ signals and $\pmb{d}_{L-U}$ will consist of the remaining $(L-U)$ signals of $\pmb{d}$. The Rayleigh quotient can now be employed to determine the optimized $\pmb{p}$:
\begin{eqnarray}\label{eq-27}
\pmb{p}=\arg\max_{\pmb{p}} \frac{\pmb{p}^H(\pmb{d}_U\pmb{d}_U^H)\pmb{p}}{\pmb{p}^H(\pmb{d}_{L-U}\pmb{d}_{L-U}^H+E[\pmb{n}\pmb{n}^H])\pmb{p}}.
\end{eqnarray}
Similar to time orthogonality,  letting $\pmb{C}=\pmb{d}_U\pmb{d}_U^H$ and $\pmb{B}= \pmb{d}_{L-U}\pmb{d}_{L-U}^H+E[\pmb{n}\pmb{n}^H]$, the optimal value of $\pmb{p}$ can be evaluated as
\begin{eqnarray}\label{eq-28}
\pmb{p}=\pmb{F}^{-1}\pmb{v},
\end{eqnarray}
where $\pmb{F}$ is the Cholesky factor of $\pmb{B}$, such that $\pmb{B}=\pmb{F}^H\pmb{F}$, and $\pmb{v}$ is the eigenvector corresponding to the maximum eigenvalue of $(\pmb{F}^H)^{-1}\pmb{C}(\pmb{F})^{-1}$. However, due to the multiplicative nature of the processing, the optimum value using (\ref{eq-28}) will produce only one signal with the maximum Rayleigh quotient. As we want to maximize $U$ observations, we must have $U$ observations coming out of the channel shortener. Therefore, we select $\pmb{v}_i$, $1\leq i \leq U$, to be the eigenvectors corresponding to the highest $U$ eigenvalues of $(\pmb{F}^H)^{-1}\pmb{C}(\pmb{F})^{-1}$. Finally, the complete preprocessing matrix $\pmb{P}$ can be given as
\begin{eqnarray}\label{eq-29}
\pmb{P}=[\pmb{p}_1,\pmb{p}_2,\cdots,\pmb{p}_U],
\end{eqnarray}
where
\begin{eqnarray}\label{eq-30}
\pmb{p}_i=\pmb{F}^{-1}\pmb{v}_i,\quad 1\leq i\leq U.
\end{eqnarray}
%
%%%%%%%%%%%%%%%%%%%%%%%%%%%%
\subsection{Proposed Optimized Channel Shortener}\label{OChannel-Shortener}
%%%%%%%%%%%%%%%%%%%%%%%%%%%%
The task of choosing a group of $U$ adds a level of optimization to the above problem as the location of $U$ signals can be anywhere within $L$. For easy of processing, as previously mentioned in~\cite{Imtiaz-2010}, these $U$ signals should be consecutively placed~\cite{Imtiaz-2012} . If we could arrange $\pmb{y}$ in a descending order such that
\begin{eqnarray}\label{eq-31}
\frac{|h_1|^2}{\sigma^2_1}\geq \frac{|h_2|^2}{\sigma^2_2} \geq \cdots \geq \frac{|h_L|^2}{\sigma^2_L},
\end{eqnarray}
then $\pmb{A}_U$ and $\pmb{d}_U$ will consist of the first $U$ strongest signals and, by applying the similar process as mentioned in section~\ref{sec-TOCS} and section~\ref{sec-FOCS}, the optimal preprocessing matrix $\pmb{P}$ can be easily carried out.

%%%%%%%%%%%%%%%%%%%%%%%%%%%%%
\section{Designing Preprocessing matrix through Rank-Reduction Techniques}\label{sec-Transform-matrixRR}
%%%%%%%%%%%%%%%%%%%%%%%%%%%%%
In this section, we propose the design of the preprocessing matrix $\pmb{P}$ with the assistance of reduced-rank techniques. In these techniques, we will utilize all the $L$ signals to design the preprocessing matrix $\pmb{P}$ instead of selecting $U$ signals out of $L$ and losing the energy of $(L-U)$ signals. We propose three different techniques, where the first two techniques are based on an eigenvalue decomposition, while the last technique utilizes the Taylor polynomial approximation (TPA).
%%%%%%%%%%%%%%%%%%%%%%%%%%%%
\subsection{Principal Component}\label{PC}
%%%%%%%%%%%%%%%%%%%%%%%%%%%%
In PC-based technique, the autocorrelation matrix $\pmb{R}$ is decomposed in terms of eigenvalues and eigenvectors~\cite{Qasim-2010,Alouini-2000,book:Dunteman}. A number of principal eigenvectors are chosen to form a detection subspace~\cite{Qasim-2010,Alouini-2000,book:Dunteman}. The decomposed autocorrelation matrix $\pmb{R}$ can be given as
\begin{eqnarray}\label{eq-32}
\pmb{R}=\pmb{\Phi}\pmb{\Lambda}\pmb{\Phi}^H=\sum_{i=1}^L \lambda_i\pmb{\phi}_i\pmb{\phi}_i^H,
\end{eqnarray}
where matrix $\pmb{\Phi}$ and matrix $\pmb{\Lambda}$ correspond to the eigenvectors and eigenvalues of $\pmb{R}$. As the autocorrelation matrix $\pmb{R}$ has distinct eigenvalues, its eigenvectors are orthonormal~\cite{book:Haykin,book:Golub}. For this reason, when selecting the largest $U$ eigenvalues, all $L$ signals are utilized instead of $U$. If these eigenvalues can be arranged in a descending order such that $\lambda_1\geq \lambda_2 \geq \cdots \geq \lambda_L$ then the eigenvectors corresponding to the first $U$ eigenvalues are retained to form the preprocessing matrix $\pmb{P}$ given as
\begin{eqnarray}\label{eq-33}
\pmb{P} = [\pmb{\phi}_1,\pmb{\phi}_2,\cdots,\pmb{\phi}_U].
\end{eqnarray}
%
%%%%%%%%%%%%%%%%%%%%%%%%%%%%
\subsection{Cross Spectral Metric}\label{CSM}
%%%%%%%%%%%%%%%%%%%%%%%%%%%%
Similar to PC, this technique utilizes the eigenvalue based technique to determine the preprocessing matrix~\cite{Goldstein-1997a,Goldstein-1997,Qasim-2010}. It has been shown in the literature that selecting the $U$ strongest eigenvalues does not necessarily represent the best set of $U$ eigenvectors, as measured by the lowest MSE~\cite{Goldstein-1997a,Goldstein-1997,Qasim-2010}. To minimize the MSE, we maximize the power of $z$, which can be represented as
\begin{eqnarray}\label{eq-34}
E_z&=&\pmb{w}^HE[\pmb{y}\pmb{y}^H]\pmb{w}=\pmb{h}^H\pmb{R}^{-1}\pmb{h}.
\end{eqnarray}
From (\ref{eq-32}), $\pmb{R}^{-1}$ can be simply written as
\begin{eqnarray}\label{eq-35}
\pmb{R}^{-1}&=&\sum_{i=1}^L \frac{\pmb{\phi}_i\pmb{\phi}_i^H}{\lambda_i}.
\end{eqnarray}
Substituting (\ref{eq-35}) into (\ref{eq-34}), we get
\begin{eqnarray}\label{eq-36}
E_z&=&\sum_{i=1}^L \frac{\pmb{h}^H\pmb{\phi}_i\pmb{\phi}_i^H\pmb{h}}{\lambda_i}
=\sum_{i=1}^L \frac{|\pmb{h}^H\pmb{\phi}_i|^2}{\lambda_i}.
\end{eqnarray}
From (\ref{eq-36}), it can be observed that for a rank of $U$, in order to maximize $E_z$, we need to collect the $U$ highest values of $|\pmb{h}^H\pmb{\phi}_i|^2/\lambda_i$. If they are arranged in a descending order such that $|\pmb{h}^H\pmb{\phi}_1|^2/\lambda_1 \geq |\pmb{h}^H\pmb{\phi}_2|^2/\lambda_2 \geq \cdots \geq |\pmb{h}^H\pmb{\phi}_L|^2/\lambda_L$ then the preprocessing matrix,
%
%\begin{eqnarray}\label{eq-36}
$\pmb{P}=[\pmb{\phi}_1,\pmb{\phi}_2,\cdots,\pmb{\phi}_U]$,
%\end{eqnarray}
%
will consist of the eigenvectors corresponding to the first $U$ values of $|\pmb{h}^H\pmb{\phi}_i|^2/\lambda_i$.

%%%%%%%%%%%%%%%%%%%%%%%%%%%%
\subsection{Taylor Polynomial Approximation}\label{TPA}
%%%%%%%%%%%%%%%%%%%%%%%%%%%%
The above two algorithms require the computation of the eigenvalues and eigenvectors which can be difficult to implement in real time applications~\cite{Burykh-2002,Honig-2000,Honig-2002,Goldstein-1998}. In some applications, computational complexity of calculating the eigenvalues will be similar to computing the inverse of the autocorrelation matrix. In such cases, eigen-decomposition based techniques cannot reduce the detection complexity. However, the Krylov subspace methods can be used to minimize the MSE, as they do not depend on the eigen-decomposition of the auto-correlation matrix $\pmb{R}$~\cite{Burykh-2002}.

Taylor Polynomial Approximation (TPA)~\cite{Qasim-2010}, Cayley-Hamilton (CH)~\cite{Burykh-2002}, power of R (POR)~\cite{Burykh-2002}, multistage Wiener Filter (MSWF)~\cite{Honig-2000,Honig-2002,Goldstein-1998}, conjugate-gradient reduced-rank filter (CGRRF)~\cite{Burykh-2002} and auxiliary vector filters (AVF)~\cite{Pados-2001,Mitra-2001}, all use Krylov subspace to design the preprocessing matrix. TPA is understood as a modified implementation of the MSWF~\cite{Honig-2000}. Furthermore, in~\cite{Burykh-2002}, TPA, POR, MSWF and CGRRF are called exact methods which were proven to be mathematically equivalent, and result in identical BER performance. It has been shown in~\cite{Mitra-2001} that AVF is equivalent to CH and MSWF. Despite the fact that all the methods are mathematically equivalent, TPA has the simplest implementation~\cite{Burykh-2002}. Therefore, TPA is only considered in this contribution.

The Taylor expansion of $\pmb{R}^{-1}$ can be expressed as
\begin{eqnarray}\label{eq-37}
\pmb{R}^{-1}&=&\mu(\mu\pmb{R})^{-1}=\mu[\pmb{I}-(\pmb{I}-\mu\pmb{R})]^{-1}\nonumber\\
&=&\mu\sum_{i=0}^\infty (\pmb{I}-\mu\pmb{R})^i,
\end{eqnarray}
where $\mu$ must satisfy $0<\mu<\lambda_{\max}$ where $\lambda_{\max}$ corresponds to maximum eigenvalue of $\pmb{R}$. Using the first $U$ values of (\ref{eq-37}), we obtain
\begin{eqnarray}\label{eq-38}
\pmb{R}^{-1}&\approx& \mu \sum_{i=0}^{U} (\pmb{I}-\mu\pmb{R})^i\nonumber\\
&=& a_0 \pmb{I} +a_1 \pmb{R}+ \cdots+a_{U-1}\pmb{R}^{U-1}.
\end{eqnarray}
The coefficients ${a_i}$ are chosen to minimize the MSE~\cite{Burykh-2002}. In the context of the TPA-assisted reduced-rank MMSE detection, the preprocessing matrix $\pmb{P}$ can be finally expressed as
\begin{eqnarray}\label{eq-39}
\pmb{P}=[\pmb{h},\pmb{R}\pmb{h},\cdots,\pmb{R}^{U-1}\pmb{h}].
\end{eqnarray}
\subsection{Remarks: Adaptive Detection}\label{Adaptivedetection}
When the destination node does not have the exact knowledge of the channel and the correlation matrix, adaptive detection can be used for reduced-rank techniques. MSWF can be implemented adaptively as mentioned in~\cite{Honig-2000,Honig-2002}. Since the exact knowledge of the correlation matrix and the cross-correlation vector is not available at the destination, MSWF suffers performance degradation, with respect to the ideal MMSE detector. In order to improve performance, joint iterative optimization (JIO) methods have been recently proposed in~\cite{Lamare-2007,Lamare-2011}. These JIO methods outperform the MSWF and the complexity depends on the choice of the adaptation algorithm. For an LMS version, the JIO method is computationally simpler than the MSWF, whereas for an RLS version, the JIO method has a cost comparable to the MSWF. Furthermore, in order to reduce the complexity of JIO methods, joint iterative interpolation, decimation, and filtering (JIDF) has been proposed in~\cite{Lamare-2009}. JIDF has a better performance as compared to JIO~\cite{Lamare-2009}. However, since we assume the destination has complete channel knowledge, it becomes unnecessary to apply adaptive techniques in this paper.

%%%%%%%%%%%%%%%%%%%%%
\section{Complexity Calculations and Analysis}\label{sec-Complexity}
%%%%%%%%%%%%%%%%%%%%%

This section demonstrates the computational complexity of each detector in a cooperative communication system. The computational complexity is measured in terms of the number of additions and multiplications required to detect a bit. The complexity of ideal MMSE detector, with a preprocessing matrix based on CS, PC, CSM and TPA techniques is summarized in Table~\ref{table:compxt}. For simplicity of calculation, only the dominant complexity terms are considered.  Also, we ignore the amount of computation required for channel estimation in the ideal MMSE detector. For the CS-based detector we assume that the selected window is always the optimal one that maximizes the SNR. Therefore, the complexity of choosing the best window is not incorporated in the analysis. Furthermore, for calculation of the computational complexity the following assumptions are employed based on~\cite{book:Golub,Martin-2005}.

\begin{itemize}

\item Multiplication of an $(M\times N)$ matrix with an $(N\times L)$ requires $M(N-1)L$ additions and $MNL$ multiplications.

\item Computing the inverse of an $(M\times M)$ matrix by using Cholesky decomposition requires $M^3/6$ additions and $M^3/6$ multiplications.

\item  Arranging the maximum eigenvalues of a matrix requires $\log_2 M$ operations where $M$ is the size of the matrix.

\end{itemize}

From Table~\ref{table:compxt}, it can be observed that the complexity of these techniques can be much lower than the ideal MMSE detector especially when $U$ is small. However, as $U$ increases, the complexity of these techniques gradually increases, and eventually exceeds that of the ideal MMSE detector. Further analysis of the complexity will be carried out in the upcoming section.

%%%%%%%%%%%%%%%%%%%%%
\section{Simulation Results and Discussion}\label{sec-results}
%%%%%%%%%%%%%%%%%%%%%
In this section, the BER performance for the proposed WSN system with $L=10$ sensors is investigated. In our simulations, the channel gains were assumed to obey the Rayleigh distribution. The transmitted signal is assumed to have unit power and the destination and all the sensors have the same noise power. All the sensors are synchronized and complete channel knowledge is available.

\subsection{BER Performance}

\begin{figure}[\protect{!htb}]
\begin{center}
\includegraphics[width=1.0\linewidth]{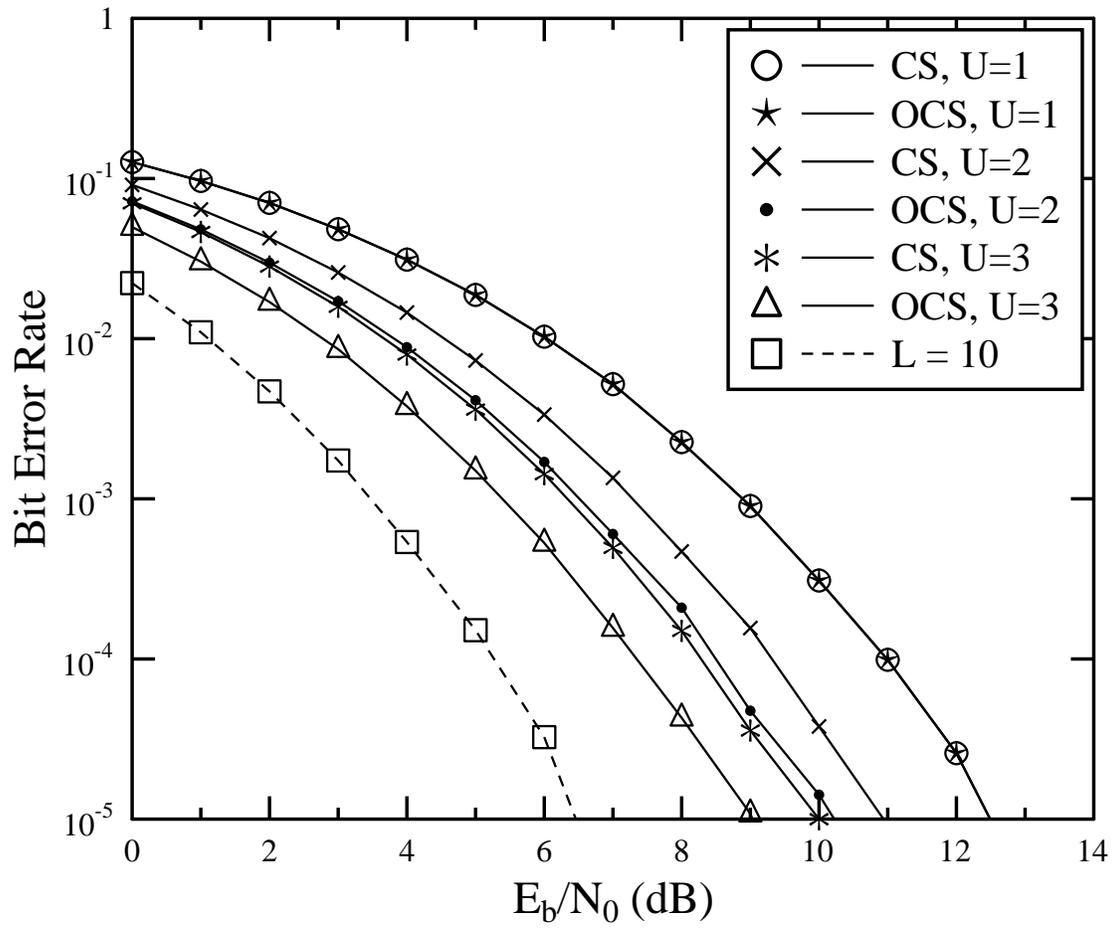}
\end{center}
\caption{{\bf Channel Shortening:} BER performance of wireless sensor networks when the sensors employ variable-gain amplification factor.}
\label{fig:CS}
\end{figure}

Fig. \ref{fig:CS} shows the BER performance as a function of the SNR per bit. The BER performance of the detection when involving all participating sensors is shown as a bench mark. It can be observed that the CS- and the optimized channel shortener (OCS)-based technique achieve the same BER performance for $U=1$. The reason of achieving the same BER performance for $U=1$ is that both schemes select the best sensor among $L$. However, for $U\geq 2$, our proposed OCS outperforms the CS by more than $1$dB. A difference of more than $3$dB can be observed as compared to all participating sensors for $U=3$. The CS- and OCS-based techniques have a similar slope as that of all participating sensors. Therefore, the CS- based, OCS-based and all participating sensors have the same diversity order. The diversity order of the system is $10$ as $L = 10$ sensors are deployed. It has already been proved in the literature that the best relay achieves full diversity order.

\begin{figure}[\protect{!htb}]
\begin{center}
\includegraphics[width=1.0\linewidth]{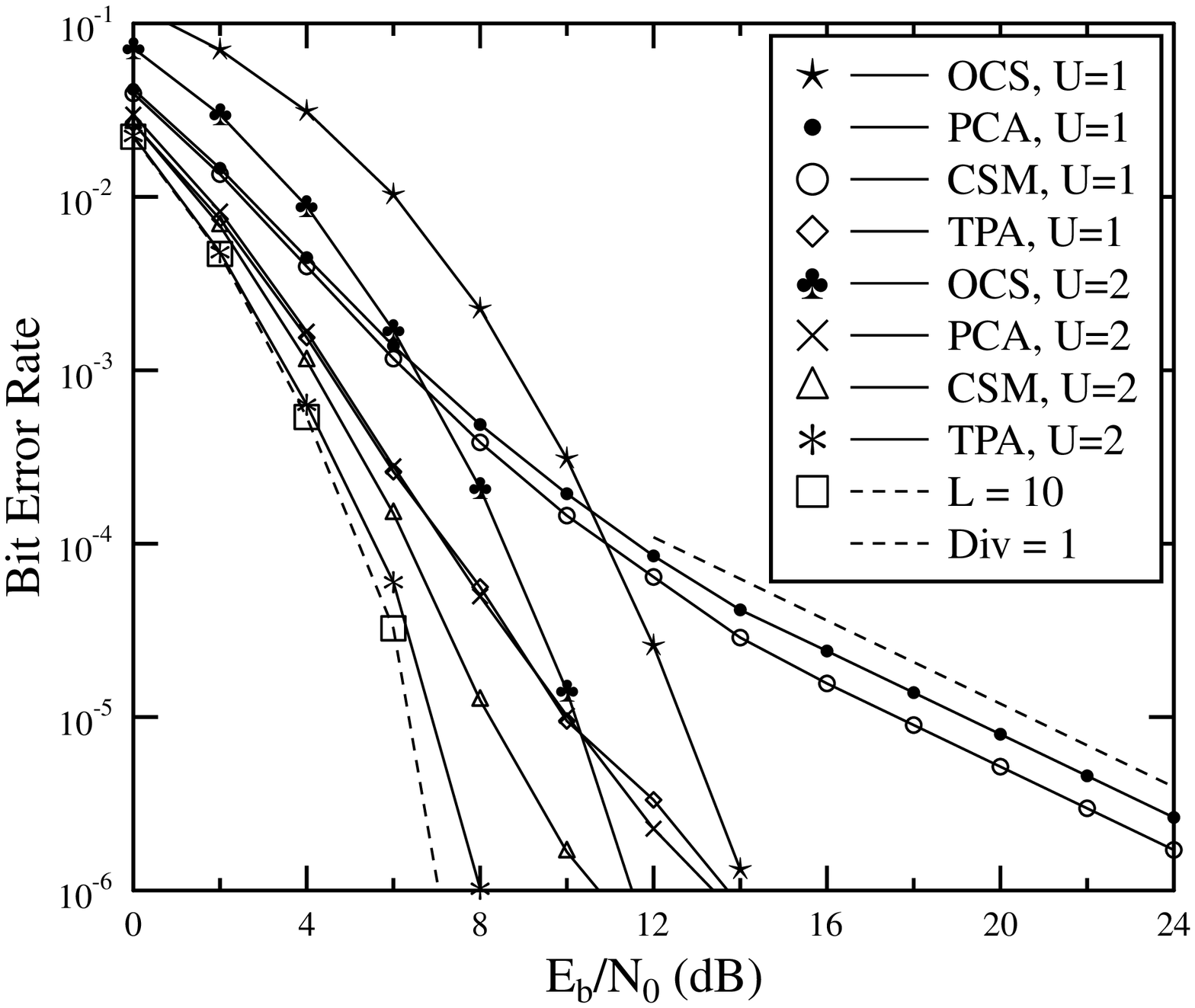}
\end{center}
\caption{{\bf Comparison:} BER performance of several detectors in wireless sensor networks when using variable-gain amplification factor.}
\label{fig:All}
\end{figure}

Fig.~\ref{fig:All} compares the BER performance when communicating with variable-gain sensors using the OCS-, PC-, CSM- and TPA-based preprocessing matrix. It can be observed that for a given size of $U$, the TPA-based technique significantly outperforms the PC-, CSM-, and OCS-assisted techniques. The PCA-based technique is the worst in terms of BER performance among the three proposed schemes. It can also be observed that OCS-based scheme outperforms the PC- and CSM-based schemes at higher SNRs. However, the BER performance of the OCS-based scheme is worse among all the considered schemes for low SNRs. It means that the PC- and CSM-based techniques lose their diversity order when reducing the size. For instance, the diversity order of PC- and CSM- based schemes for $U=1$ matches our analytic result which is addressed in the appendix.

Finally, in Fig.~\ref{fig:fixedgainrelay}, we compare the BER performance of WSNs when using fixed-gain amplification factor. It can be observed that the TPA-based scheme is equivalent to all participating sensors for $U=1$. TPA-based technique also performs better than the PC-, CSM- and OCS-assisted schemes. The PC-based scheme is equivalent to the CSM-assisted scheme in terms of BER performance. It can also be observed that the BER performances of all these schemes are much superior to the OCS-based scheme. Moreover, the slopes for all schemes are the same, therefore, the system will have a diversity order of $L=10$. We have tried to address this diversity order for PC- and CSM-based scheme in the appendix.
\subsection{Complexity Analysis}
In Fig.~\ref{fig:complex}, the number of operations is plotted with respect to selected size of $U$. It can be observed that, as $U$ increases, more operations are required to detect a bit. The complexity of the CS-based scheme increases quadratically and is more than that of the TPA-based scheme. The complexity of the PC- and CSM-based schemes is similar. For high $U$, the complexity of PC- and CSM-based techniques is lower than that of the CS- and TPA-based techniques.

Finally, Table~\ref{compxt-2} presents the computation saving versus difference in SNR as compared to the ideal MMSE detector at BER of $10^{-4}$. Firstly, increasing the value of $U$ decreases the computational saving and the difference in SNR. Secondly, for $U=3$, OCS-based preprocessing matrix neither gives any computational saving nor the difference in SNR gets close to the ideal MMSE detector. Therefore, if the selection is more than $3$ sensors, it is better to deploy ideal MMSE detection as compared to the OCS-based detector. Thirdly, the complexity saving of PC-based preprocessing matrix becomes larger than that of CSM-based preprocessing matrix, although this advantage in complexity saving results in more difference in SNR as compared to ideal MMSE detector. Finally, it can be concluded that designing TPA-based preprocessing matrix gives us more computational saving as well as less difference in SNR as compared to all the other considered techniques.

\section{Conclusions}\label{sec-summary}
In this paper, channel-shortening (CS) and rank-reduction techniques have been proposed for cooperative WSNs to reduce the complexity of the ideal MMSE detector while maintaining the BER performance. The proposed CS technique outperforms the previously known CS techniques by more than $1$dB when $U\geq 2$. The performance and complexity of the proposed reduced-rank techniques are superior to the CS technique when deploying fixed-gain amplification factor. However, a tradeoff can be observed between the complexity and BER performance when the sensors utilize variable-gain amplification factor. The cross spectral metric (CSM)-based technique outperforms the principal component (PC)-based technique in terms of BER, but with a modest increase in complexity. While, the Taylor polynomial approximation (TPA)-based technique reaches the same BER performance as the ideal MMSE for $U=3$. Our future research will concentrate on implementing adaptive rank-reduction schemes when the channel knowledge is not available.

%Finally, it can be concluded that for a wireless sensor network where there are a large number of sensors, these reduced rank techniques can be employed which can achieve a reasonable BER performance but with low complexity.  As our proposed techniques will reduce the computational complexity of the system, it will result in lower resources usage as compared to ideal MMSE.

\appendix\label{Eigen-value Calculations}

In this appendix, we carry out a case study for the diversity order of PCA- and CSM-based rank reduction using $U=1$. The correlation matrix can be given as $\pmb{R}=\pmb{h}\pmb{h}^H+\pmb{\Sigma}$ and we can asymptotically approximate $\pmb{R}\doteq \pmb{h}\pmb{h}^H$ as the SNR goes to infinity (i.e., $\sigma_1^2,\sigma_2^2\rightarrow 0$), where $\doteq$ denotes asymptotic value in high SNR. Then, the eigenvector corresponding to the maximum eigenvalue is approximately $\pmb{\phi}_1\doteq\pmb{h}$ which is the preprocessing matrix $\pmb{P}$ when $U=1$ in PCA-based reduced rank technique (it is also a preprocessing vector in TPA-based technique with $U=1$.). Multiplying this preprocessing vector by the received signal, the output signal can be given as
\begin{eqnarray}~\label{eq-40}
\pmb{P}^H\pmb{y}=\pmb{\phi}_1^H\pmb{y}=\pmb{h}^H(\pmb{h}b+\pmb{n}),
\end{eqnarray}
and the received SNR can be computed as
\begin{eqnarray}~\label{eq-41}
\gamma\doteq\frac{\left(\sum_{\ell=1}^L |h_\ell|^2\right)^2}{\sum_{\ell}^L \sigma_\ell^2|h_\ell|^2}.
\end{eqnarray}
The diversity order of the received SNR can be defined as~\cite{Xu-2010}
\begin{eqnarray}~\label{eq-42}
d =\lim_{\rho\rightarrow \infty} -\frac{\log(F_{\gamma}(x))}{\log \rho},
\end{eqnarray}
where $\rho$ denotes the SNR and $F_{\gamma}(x)$ is the cumulative distribution function (CDF) of $\gamma$. Without loss of generality, we set $\rho=\frac{1}{\sigma^2}$ by assuming $\sigma_{R_\ell}^2=\sigma_{D_\ell}^2=\sigma^2$ and $E_S=E_{R_\ell}=1$ for all $\ell=1,2,\ldots,L$. Then, let us look at the diversity order for fixed- and variable-gain amplifications. Throughout this appendix, the diversity gain is calculated with the help of the CDF of the received SNR which is equivalent to the outage probability of the received SNR.

\subsection{Fixed-Gain Amplification}

Applying fixed-gain amplification factor in high SNR, i.e., $\zeta_{R_\ell}^2\doteq 1$, and substituting it and $\sigma_\ell^2 \doteq \sigma^2(|h_{R_\ell D}|^2+1)$ into (\ref{eq-41}) yields
\begin{eqnarray}~\label{eq-43}
\gamma_{f} \doteq \frac{\left(\sum_{\ell=1}^L |h_{SR_\ell}|^2|h_{R_\ell D}|^2\right)^2}{\sigma^2\sum_{\ell=1}^L|h_{SR_\ell}|^2|h_{R_\ell D}|^2(|h_{R_\ell D}|^2+1)}.
\end{eqnarray}
It is also intractable to compute the exact CDF of $\gamma_f$ and we first find the tractable lower bound of $\gamma_f$. The lower bound on the diversity order in terms of SNR will emphasize that the system will have at least the diversity order of lower bound of $\gamma_f$. Intuitively, $\gamma_f$ is bounded by $\gamma_{f,lb}$ which is represented as
\begin{eqnarray}~\label{eq-44}
\gamma_f \!&\!>\!&\! \frac{\sum_{\ell=1}^L |h_{SR_\ell}|^2|h_{R_\ell D}|^2}{\sigma^2\max_{\ell}|h_{R_\ell D}|^2+1}\nonumber\\
\!&\!>\!&\!\frac{\min_\ell |h_{R_\ell D}|^2}{\max_\ell |h_{R_\ell D}|^2 +1 }\cdot\sum_{\ell=1}^L |h_{SR_\ell}|^2\rho \triangleq \gamma_{f,lb}.
\end{eqnarray}
Defining two random variables, $X=\frac{\min_\ell |h_{R_\ell D}|^2}{\max_\ell |h_{R_\ell D}|^2 +1 }$ and $Y=\sum_{\ell=1}^L |h_{SR_\ell}|^2\rho$, we can easily observe that the random variable $X$ is irrelevant of $\rho$ and $Y$ is the central chi-square random variable with $2L$ degrees of freedom with probability density function (PDF)~\cite{book:alouini}
\begin{eqnarray}~\label{eq-45}
f_Y(y) = \frac{1}{(L-1)!\rho^L}y^{L-1}e^{-y/\rho}.
\end{eqnarray}
By substituting the Taylor expansion of $e^{-y/\rho}=\sum_{\ell=0}^\infty(-y/\rho)^\ell$ into the above PDF, the CDF can be represented as
\begin{eqnarray}~\label{eq-46}
F_Y(y) = \frac{y^L}{L!}\rho^{-L}+\emph{o}(\rho^{-L}).
\end{eqnarray}
The CDF of $\gamma_{f,lb}$ can be represented as
\begin{eqnarray}~\label{eq-47}
F_{\gamma_{f,lb}}(x)\!&\! =\! &\!E_X\left[F_Y(x/X)\right]\nonumber\\
\!&\! =\! &\! \frac{x^L}{L!}E\left[\frac{1}{X^L}\right]\rho^{-L}+\emph{o}(\rho^{-L}).
\end{eqnarray}
Finally, the diversity order of $\gamma_{f}$ is given by
\begin{eqnarray}~\label{eq-48}
d_f \geq \lim_{\rho\rightarrow \infty} -\frac{\log\left( \rho^{-L}\left(\frac{x^L}{L!}E\left[\frac{1}{X^L}\right]+\emph{o}(1)\right)\right )}{\log \rho}=L.
\end{eqnarray}
%we have to rephrase
It is straightforward to achieve a diversity order of $L$ at maximum with $L$ independent channels and therefore we conclude that $d_f=L$. Hence, full diversity is achieved when employing fixed gain amplification even with $U=1$.

\subsection{Variable-Gain Amplification}
Applying variable-gain amplification factor in high SNR, i.e., $\zeta_{R_\ell}^2\doteq 1/|h_{SR_\ell}|^2$, and substituting it and $\sigma_\ell^2\doteq \sigma^2(|h_{R_\ell D}|^2/|h_{SR_\ell}|^2+1)$ into (\ref{eq-41}) yields
\begin{eqnarray}~\label{eq-49}
\gamma_{v} \doteq \frac{\left(\sum_{\ell=1}^L |h_{R_\ell D}|^2\right)^2}{\sigma^2\sum_{\ell=1}^L|h_{R_\ell D}|^2(|h_{R_\ell D}|^2/|h_{SR_\ell}|^2+1)}.
\end{eqnarray}
Due to the intractability of the exact CDF of $\gamma_v$, we should find the upper-bound of $\gamma_v$ which does not affect the diversity order of $\gamma_v$. By omitting some parts of denominator, $\gamma_v$ is upper-bounded such that
\begin{eqnarray}~\label{eq-50}
\gamma_v < \frac{\left(\sum_{\ell=1}^L |h_{R_\ell D}|^2\right)^2}{\sigma^2\sum_{\ell=1}^L|h_{R_\ell D}|^4/|h_{SR_\ell}|^2}.
\end{eqnarray}
Using the inequality, \mbox{$\sum_{\ell=1}^L|h_{R_\ell D}|^4/|h_{SR_\ell}|^2 > |h_{R_k D}|^4/|h_{SR_k}|^2$} for any $k$, $\gamma_v$ is further bounded as
\begin{eqnarray}~\label{eq-51}
\gamma_v < \min_\ell  \left(\sum_{k=1}^L |h_{R_k D}|^2/|h_{R_\ell D}|^2\right)^2\cdot|h_{SR_\ell}|^2\rho =\gamma_{v,ub}.
\end{eqnarray}
Defining $X_\ell=\left(\sum_{k=1}^L |h_{R_k D}|^2/|h_{R_\ell D}|^2\right)^2$ and $Y_\ell = |h_{SR_\ell}|^2\rho$ temporarily, we can say that $X_\ell$ is irrelevant of the SNR $\rho$ and $Y_\ell$ is exponentially distributed with $Y_\ell\sim \chi_2^2 (\rho)$. The CDF of $\gamma_{v,ub}$ can be represented as
\begin{eqnarray}~\label{eq-52}
F_{\gamma_{v,ub}}(x)\!&\!=\!&\!\sum_{\ell=1}^L\mathrm{Pr}\Big[X_\ell Y_\ell < x | X_\ell Y_\ell < \min_{k\neq \ell} X_k Y_k\Big]\nonumber\\
&\times&\mathrm{Pr}\Big[X_\ell Y_\ell < \min_{k\neq \ell} X_k Y_k\Big]\nonumber\\
\!&\!=\!&\! \mathrm{Pr}\Big[X_\ell Y_\ell < x \Big| X_\ell Y_\ell < \min_{k\neq \ell} X_k Y_k\Big],\nonumber\\
\!&\!>\!&\! \mathrm{Pr}[X_\ell Y_\ell < x],\label{CDFlb}
\end{eqnarray}
where the second equality holds because both probabilities in summation are equally probable over $\ell=1,2,\ldots,L$. This conditional CDF is cumbersome to calculate because of the correlation between random variables in order over $\ell$. By omitting the condition, the third inequality holds. By using the Taylor expansion of the CDF of $Y_\ell$, $F_{Y_\ell}(y)=1-e^{-y/\rho}=-\sum_{k=1}^\infty (-y/\rho)^k$, the probability in (\ref{CDFlb}) is represented as
\begin{eqnarray}~\label{eq-53}
\mathrm{Pr}[X_\ell Y_\ell < x] \!&\!=\!&\! E_{X_\ell}\Big[F_{Y_\ell}(x/X_\ell)\Big]\nonumber\\
\!&\!=\!&\! E\left[\frac{x}{X_\ell}\right]\rho^{-1}+\emph{o}(\rho^{-1}).
\end{eqnarray}
Finally, the diversity order of $\gamma_v$ is given by
\begin{eqnarray}~\label{eq-54}
d_v&\leq& \lim_{\rho\rightarrow \infty}-\frac{\log(F_{\gamma_{v,ub}}(x))}{\log \rho}\nonumber\\
&\leq& \lim_{\rho\rightarrow \infty}-\frac{\log\left(\rho^{-1}\Big(E\Big[\frac{x}{X_\ell}\Big]+\emph{o}(1)\Big)\right)}{\log \rho}=1.
\end{eqnarray}
It is straightforward to achieve at least the diversity order of $1$ like a single channel scenario and therefore we say $d_v=1$ here. When variable-gain amplification factor is used in case that $U=1$, a loss in diversity order can be observed as compared to fixed-gain amplification factor which was discussed in the previous section.
%
%%%%%%%%%%%%%%%%%%%%%%%%%%%%%%%%%%%%%%%%%%%%%%%%%%%%%%%%%%%%%%%%%%%%%%%%%

%
\begin{table*}[!hbt]
\caption{Computational complexity.}\label{table:compxt}
\begin{center}
\begin{tabular}{|l|c|c|}
\hline
&\multicolumn{2}{c|}{Number of operations per symbol}\\
\cline{2-3}\raisebox{3pt}[0pt][0pt]{Algorithm}&Additions & Multiplications\\
\hline
Ideal MMSE detector&$L^3/3+L^2+L$&$L^3/3+L^2+2L$\\
\hline
CS-based MMSE detector&$3U^2L+17U^3/3+LU+\log_2U-L-3U$&$3U^2L+17U^3/3+5U^2+UL+U$\\
\hline
PC-based MMSE detector&$L^3/6+U^3/3+2UL+U^2+U\log_2 L+L+U$&$L^3/6+U^3/3+2UL+U^2+L+U$\\
\hline
CSM-based MMSE detector&$L^3/6+U^3/3+2UL+U^2+U\log_2 L+2L+U$&$L^3/6+U^3/3+2UL+U^2+4L+U$\\
\hline
TPA-based MMSE detector&$U^3/3+(U-1)L^2+2UL+U^2+L$&$U^3/3+(U-1)L^2+2UL+U^2+L$\\
\hline
\end{tabular}
\end{center}
\end{table*}

\begin{table}[!ht]
\caption{Computation Saved vs Difference in SNR as compared to the Ideal MMSE detector at BER of $10^{-4}$.}\label{compxt-2}
\begin{center}
\begin{tabular}{|c|c|c|c|c|}
\hline
\multicolumn{2}{|c|}{Preprocessing}&Computation&Difference&Difference\\
\multicolumn{2}{|c|}{Based}&Saved& in SNR& in SNR\\
\multicolumn{2}{|c|}{Detector}&w.r.t ideal& with fixed&with variable\\
\multicolumn{2}{|c|}{}&MMSE&gain sensors &gain sensors\\
\hline
&U=1 &\quad 91.05\% &4.85 &5.6\\
\cline{3-5}
OCS& U=2&\quad 59.66\%&2.25&3.3\\
\cline{3-5}
&U=3&\quad 0\%&1.25&2\\
\hline
&U=1&54.74\%&0.25&6.3\\
\cline{3-5}
PC&U=2&48.42\%&0.0&1.8\\
\cline{3-5}
&U=3&40.76\%&0.0&0.8\\
\hline
&U=1&50.23\%&0.05&4.85\\
\cline{3-5}
CSM &U=2&43.91\%&0.0&1.3\\
\cline{3-5}
&U=3&36.24\%&0.0&0.6\\
\hline
&U=1&92.93\%&0.0&1.9\\
\cline{3-5}
TPA&U=2&64.66\%&0.0&0.3\\
\cline{3-5}
&U=3&35.04\%&0.0&0.0\\
\hline
\end{tabular}
\end{center}
\vspace{-0.25cm}
\end{table}

\begin{figure}[\protect{!htb}]
\begin{center}
\includegraphics[width=1.0\linewidth]{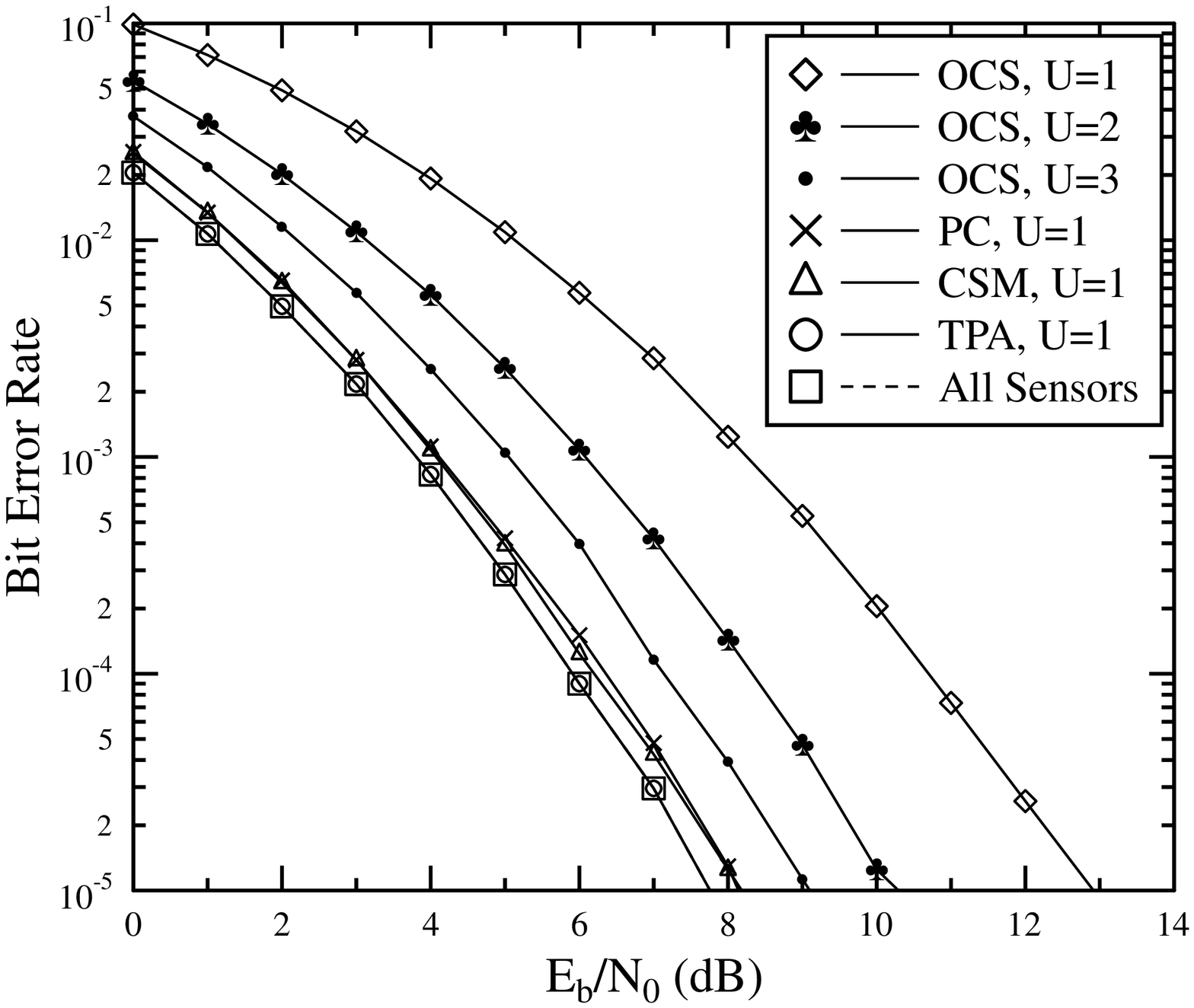}
\end{center}
\caption{{\bf Comparison:} BER performance of several detectors in wireless sensor networks when using fixed-gain amplification factor.}
\label{fig:fixedgainrelay}
\end{figure}
\begin{figure}[\protect{!htb}]
\begin{center}
\includegraphics[width=1.0\linewidth]{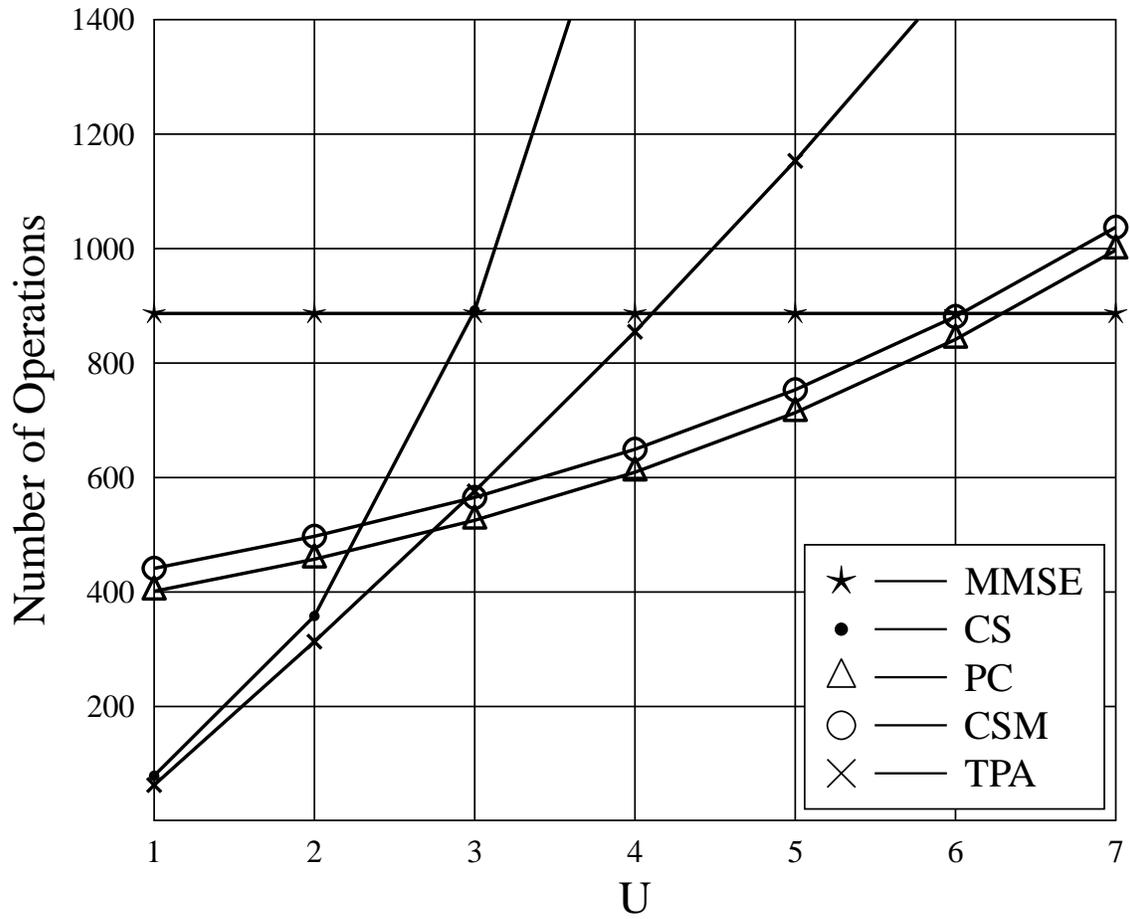}
\end{center}
\vspace{-0.6cm}
\caption{Number of operations versus the selected size $U$.}
\label{fig:complex}
\end{figure}

\end{document}